%% file: main.tex
\input{_0_vars}

\ifdefined\isarxiv
\documentclass[11pt]{article}
\usepackage[numbers]{natbib}
\else
\documentclass[sigconf,anonymous,review]{acmart}
\AtBeginDocument{%
  }
\setcopyright{acmlicensed}
\copyrightyear{2025}
\acmYear{2025}
\acmConference[KDD '26]{ACM KDD }{August 09--13,
  2026}{Jeju, Korea}
\fi

\ifdefined\isarxiv

\usepackage{amsmath}
\usepackage{amsthm}
\usepackage{amssymb}
\usepackage{algorithm}
\usepackage{subfig}
\usepackage{algpseudocode}
\usepackage{graphicx}
\usepackage{grffile}
\usepackage{wrapfig,epsfig}
\usepackage{url}
\usepackage{xcolor}
\usepackage{epstopdf}

\usepackage{bbm}
\usepackage{dsfont}

\else

\fi

\ifdefined\isarxiv

\else
\usepackage{amsmath}
\usepackage{amsthm}
\usepackage{algorithm}
\usepackage{subfig}
\usepackage{algpseudocode}
\usepackage{grffile}
\usepackage{wrapfig,epsfig}
\usepackage{url}
\usepackage{xcolor}
\usepackage{epstopdf}

\usepackage{bbm}
\usepackage{dsfont}
\fi
 
\allowdisplaybreaks

\ifdefined\isarxiv

\usepackage{tikz}
\usepackage{hyperref}  
\hypersetup{colorlinks=true,citecolor=blue,linkcolor=blue} 
\usetikzlibrary{arrows}
\usepackage[margin=1in]{geometry}

\else
\fi
 
\graphicspath{{./figs/}}

\theoremstyle{plain}
\newtheorem{theorem}{Theorem}[section]

\newtheorem{definition}[theorem]{Definition}

\input{_1_maths}

\begin{document}

\ifdefined\isarxiv

\date{}
\title{\paperTitle}
\author{\paperAuthor}

\else
\title{\paperTitle}

\begin{abstract}
\input{00_abstract}
\end{abstract}





\maketitle

\fi

\ifdefined\isarxiv
\begin{titlepage}
  \maketitle
  \begin{abstract}

\input{00_abstract}
  \end{abstract}
  \thispagestyle{empty}
\end{titlepage}

\newpage

\else

\fi


\input{_2_body}

\ifdefined\isarxiv
\bibliographystyle{alpha}
\bibliography{ref}
\else
\bibliographystyle{ACM-Reference-Format}
\bibliography{ref}
\fi





\end{document}

%% file: _0_vars.tex
\def\isarxiv{1}

\def\paperTitle{When a Paper Has 1000 Authors: Rethinking Citation Metrics in the Era of LLMs}

\def\paperAuthor{
Weihang Guo\thanks{\texttt{weihang.guo2001@gmail.com}.}
\and
Zhao Song\thanks{\texttt{magic.linuxkde@gmail.com}. University of California, Berkeley.}
\and 
Jiahao Zhang\thanks{\texttt{ml.jiahaozhang02@gmail.com}.}
}



%% file: _1_maths.tex
\usepackage{booktabs} 

\newcommand{\wt}{\widetilde}

\renewcommand{\tilde}{\wt}

%% file: 00_abstract.tex
Author‐level citation metrics provide a practical, interpretable, and scalable signal of scholarly influence in a complex research ecosystem. It has been widely used as a proxy in hiring decisions. However, the past five years have seen the rapid emergence of large-scale publications in the field of large language models and foundation models, with papers featuring hundreds to thousands of co-authors and receiving tens of thousands of citations within months. For example, Gemini~\cite{team2023gemini} has 1361 authors and has been cited around 4600 times in 19 months. In such cases, traditional metrics, such as total citation count and the $h$-index, fail to meaningfully distinguish individual contributions. Therefore, we propose the following research question: How can one identify standout researchers among thousands of co-authors in large-scale LLM papers? This question is particularly important in scenarios such as academic hiring and funding decisions.
In this paper, we introduce a novel citation metric designed to address this challenge by balancing contributions across large-scale and small-scale publications. We propose the SBCI index, analyze its theoretical properties, and evaluate its behavior on synthetic publication datasets. Our results demonstrate that the proposed metric provides a more robust and discriminative assessment of individual scholarly impact in the era of large-scale collaborations.

%% file: _2_body.tex

\input{01_intro}

\input{02_preli}
\input{03_related}
\input{04_problem}
\input{05_methods}
\input{06_experiment}
\input{07_discussion}

\input{08_conclusion}

%% file: 01_intro.tex
\section{Introduction}
Citations~\cite{hir05} serve as a proxy for influence. They provide measurable evidence that a piece of work has contributed to the progress of scientific knowledge, been recognized by peers, and influenced subsequent research~\cite{mmd23}. With the exponential growth of scientific output across disciplines, citation-based metrics have become essential tools for evaluating individual researchers, guiding funding decisions, determining academic promotions, and shaping institutional rankings~\cite{apf23, hww+15}. However, while citation metrics aim to quantify research impact objectively, their design often reflects assumptions about authorship structure and citation behavior that may no longer hold in today’s research landscape~\cite{ein+24, khf+13}, particularly in fields dominated by large-scale collaborations such as Large Language Models~(LLMs)~\cite{team2024gemini, yang2025qwen3}, Large Vision Models~(LVMs)~\cite{gunter2024apple, sun2024hunyuan}, and Video Generation Models~(VGM)~\cite{polyak2024movie, peng2025open}.

\begin{table*}[!ht]
\centering
\begin{tabular}{lccccccccc}
\toprule
Model & Date & Months & Total & Core & Additional & Total \\
name & posted & posted & citations & authors & authors & authors \\
\midrule
Gemini~\cite{team2023gemini} & Dec 23 & 19 & 4612 & 714 & 647 & 1361 \\
Gemini 1.5~\cite{team2024gemini} & Mar 24 & 16 & 2032 & 689 & 449 & 1138 \\
LLaMA 3~\cite{grattafiori2024llama3} & Jul 24 & 12 & 5411 & 236 & 323 & 559 \\
DeepSeek R1~\cite{guo2025deepseekr1} & Jan 25 & 6 & 1985 & 18 & 176 & 194 \\
Qwen 3~\cite{yang2025qwen3} & May 25 & 2 & 1952 & 60 & 118 & 178 \\
DeepSeek V2~\cite{liu2024deepseekv2} & May 24 & 14 & 320 & 157 & 0 & 157 \\
Apple Intelligence~\cite{gunter2024apple} & Jul 24 & 12 & 67 & 155 & 0 & 155 \\
DeepSeek V3~\cite{liu2024deepseekv3} & Dec 24 & 7 & 1032 & 143 & 0 & 143 \\
Hunyuan~\cite{sun2024hunyuan} & Nov 24 & 8 & 29 & 108 & 0 & 108 \\
Seedance 1.0~\cite{gao2025seedance} & Jun 25 & 1 & 0 & 44 & 49 & 93 \\
MovieGen~\cite{polyak2024movie} & Oct 24 & 9 & 6 & 56 & 32 & 88 \\
LLaMA 2~\cite{touvron2023llama2} & Jul 23 & 24 & 16022 & 2 & 66 & 68 \\
Wan~\cite{wan2025wan} & Mar 25 & 4 & 28 & 61 & 0 & 61 \\
Qwen 1~\cite{bai2023qwen} & Sep 23 & 22 & 3738 & 47 & 0 & 47 \\
Apple MM1~\cite{mckinzie2024mm1} & Mar 24 & 16 & 303 & 7 & 25 & 32 \\
OpenSora 2.0~\cite{peng2025open} & Mar 25 & 4 & 15 & 17 & 14 & 31 \\
LLaMA 1~\cite{touvron2023llama1} & Feb 23 & 29 & 17329 & 5 & 9 & 14 \\
\bottomrule
\end{tabular}

\caption{Model authorship and citation statistics. The \emph{Months posted} column shows how many months have elapsed since each paper was posted on arXiv, and the \emph{Total citations} column gives the number of citations accumulated up to \emph{July 1, 2025}.}\label{tab:model_stats}
\end{table*}

Over the past five years, natural language processing has undergone an extraordinary transformation driven by large-language-model research. Breakthrough systems such as GPT-3~\cite{brown2020gpt3} and their successors have pushed model sizes from the tens of millions of parameters to well over one hundred billion. Tasks such as data collection, pretraining, fine-tuning, and large-scale evaluation that were once tractable on a single GPU now demand sprawling compute clusters, bespoke software stacks, and multidisciplinary engineering teams~\cite{zrg22}. Coordinating these efforts requires advanced project management practices, including versioned datasets, distributed training pipelines, and rigorous reproducibility checks~\cite{bbm24}. Consequently, the research of foundation-model development has shifted decisively toward industry laboratories with the capital to underwrite both the hardware and human resources involved. The resulting technical reports routinely consists of hundreds or even thousands of author names, and these publications often accumulate thousands of citations within mere months of release, with the most prominent papers surpassing tens of thousands of citations in a single year.

Historically, research impact has been quantified primarily through metrics designed at the author level, among which total citation count, the $h$-index~\cite{hir05}, and $i_{10}$-index are most widely used. These metrics offer a simple, interpretable summary of an individual’s academic influence based on how often their work is cited by others~\cite{bd07}. However, these metrics rest on historical assumptions that no longer match today’s scientific landscape. At the time they were developed, most papers had a similar number of authors, and citation counts typically ranged from a few dozen to a few hundred per paper. Conversely, in the era of hyper-authored LLM research, a single paper with hundreds of co-authors and tens of thousands of citations can dramatically inflate individual metrics. This makes it difficult to compare authors who have contributed to such blockbuster papers with those who have not. Even among contributors to the same hyper-authored, hyper-cited work, traditional citation indices offer little resolution to distinguish each author. Given the following four candidates:
\begin{itemize}
  \item \emph{Candidate A:} Two hyper-authored LLM papers, no other publications.
  \item \emph{Candidate B:} Four hyper-authored LLM papers, no other publications.
  \item \emph{Candidate C:} Two hyper-authored LLM papers plus four top-conference ML papers.
  \item \emph{Candidate D:} Ten top-conference ML papers, no hyper-authored LLM papers.
\end{itemize}
Which metric should be used to distinguish them: total citations, $h$-index, number of papers, or some other measure? In this paper, we mathematically formalize this scenario and introduce a novel metric that balances credit among hyper-authored consortium papers, small-team research articles, highly cited industry technical reports, and traditional academic machine-learning publications. Our objective is to equip anyone outside the LLM, VLM, or VGM domain with a helpful, practical, interpretable metric that accurately reflects an author’s research ability and scholarly impact across both hyper-authored and small-team publications.

\textbf{Roadmap:} We begin in Section~\ref{sec:related_work} by surveying widely used citation metrics including the $i_{10}$-index, $h$-index, $g$-index and their coauthorship‐adjusted variants. We highlight how industry‐driven foundation models and hyperauthorship violate their core assumptions.  Section~\ref{sec:problem_definition} then formalizes our problem setting and the fundamental properties any citation metric must satisfy.  In Section~\ref{sec:methods}, we introduce two new measures, the Exponential Fractional $h$-Index and the Scale‐Balanced Citation Index. We prove that they meet the desired properties and describe a grid‐search procedure for tuning SBCI’s hyperparameters by optimizing a composite objective balancing discriminative power, statistical balance, and ranking stability.  Section~\ref{sec:experiment} presents a synthetic dataset designed to reflect real‐world citation and collaboration patterns, reports the results of our parameter search, and demonstrates SBCI’s behavior alongside traditional metrics in a six‐candidate case study.  We discuss the practical implications and directions for future work in Section~\ref{sec:discussion} and conclude the paper in Section~\ref{sec:conslusion}.

%% file: 02_preli.tex

%% file: 03_related.tex
\section{Related Work}\label{sec:related_work}
In this section, we review prior work on citation metrics and the evolving structure of research teams in the foundation model era.  Section~\ref{sec:related_work_index} formalizes the most widely used author‐level indices, such as the $i_{10}$-index, $h$-index, $g$-index, and their coauthorship‐adjusted variants, and highlights their underlying assumptions.  Section~\ref{sec:related_work_fm} then examines how the shift of large‐scale AI development from academia to industry has driven unprecedented team sizes and compute requirements in foundation model research.  Finally, Section~\ref{sec:related_work_team} surveys efforts to define “large” versus “small” collaboration thresholds.  

\subsection{Formal Definitions of Popular Metrics}\label{sec:related_work_index}
In this subsection, we present formal definitions for the most widely used citation-based metrics. We begin with the index currently adopted by Google Scholar~\cite{google12}: $i_{10}$ and $h$-index. The $i_{10}$ index counts the number of an author’s publications that have received at least ten citations. 
\begin{definition}[$i_{10}$-index]
    Let an author have $N$ papers where paper $i$ has $c_i$ citations. Then the $i_{10}$-index is
    \begin{align*}
        i_{10} := \sum^{N}_{i=1}\mathbf{1}_{\{c_i \geq 10\}}, 
    \end{align*}
    where $\mathbf{1}_{\{\cdot\}}$ is the indicator function. 
\end{definition}

Notice that for any $a \in \mathbb{N}^{+}$, the metric $i_{a}$
counts the number of an author’s publications cited at least $a$ times. The reason the $i_{10}$-index has become the most popular of this family rests on the following assumption:


$h$-index~\cite{hir05} measure the largest $h$ such that the author has at least $h$ publications with at least $h$ citations.  Essentially, it indicates both the productivity and citation impact of the publications. 
\begin{definition}[$h$-index from~\cite{hir05}]\label{def:hindex}
    Let an author have $N$ papers with citation counts $c_1, \cdots, c_N$ sorted so that $c_1\geq c_2 \geq \cdots \geq c_N$. Then the $h$-index is 
    \begin{align*}
        h:=\max\{ i\in \mathbb{N} \mid c_i \geq i\}. 
    \end{align*}
\end{definition}

The $h$-index provides a fair estimate under the following assumptions:

The $g$-index~\cite{egg06} is proposed to address the $h$-index’s insensitivity to a small number of exceptionally highly cited papers.

\begin{definition}[$g$-index from~\cite{egg06}]
    Let an author have $N$ papers with citation counts $c_1, \cdots, c_N$ sorted so that $c_1\geq c_2 \geq \cdots \geq c_N$. The the $g$-index is 
    \begin{align*}
        g:=\max\{ k\in \mathbb{N} \mid \sum^{k}_{i=1}c_i \geq k^2\}
    \end{align*}
\end{definition}

The individual $h$-index is proposed to correct for coauthorship by dividing the standard $h$-index by the average number of authors, enabling fairer comparisons among researchers with different collaboration patterns.

\begin{definition}[Individual $h$-index from~\cite{bck06}]
    Let an author have $N$ papers with citation counts $c_1, \cdots, c_N$ sorted so that $c_1\geq c_2 \geq \cdots \geq c_N$. Let $h=\max\{ i\in \mathbb{N}~|~ c_i \geq i\}$ be the $h$-index in Definition~\ref{def:hindex}. Let $P_h=\{p_1, \cdots, p_h\}$ be  the set of those $h$ papers and $a_i$ be the number of authors on paper $p_i$. Define the total author count as $N_a := \sum^{h}_{i=1}a_i$. Then the individual $h$-index is 
    \begin{align*}
        h_{\rm I} := \frac{h^2}{N_a}. 
    \end{align*}
\end{definition}

The fractional $h$-index is proposed to adjust for coauthorship by assigning each paper’s citations equally among its authors before computing the index.

\begin{definition}[Fractional $h$-index from~\cite{egg08}]
    Let an author have $N$ papers where paper $i$ has $c_i$ citations and $a_i$ authors. Define the fractional citation count for paper $i$ as $f_i := \frac{c_i}{a_i}$. Sort the papers so that $f_1 \geq f_2 \geq \cdots \geq f_N$. Then the fractional $h$-index is 
    \begin{align*}
        h_{\rm frac} := \max\{ i\in \mathbb{N} \;|\; f_i\geq i\}
    \end{align*}
\end{definition}


\subsection{Industry Lead in Foundation Model Research}\label{sec:related_work_fm}
Since the early 2010s, access to the massive compute resources needed for state-of-the-art AI research has largely migrated from universities to industry~\cite{shh22}. The rise of large self-supervised models, trained on enormous unlabeled datasets, has made computational capacity the central constraint in advancing machine learning~\cite{bha21}. Over the last decade, the scale of compute required for training has exploded by factors ranging from hundreds of thousands to tens of millions~\cite{shh22,ghl22} and driven costs for models into the tens of millions of dollars. This surge has rendered the development of even widely adopted architectures, such as medium and large scale BERT models, financially out of reach for most academic labs~\cite{sps20,ibl21}.

As compute requirements surge, the process of building and training modern machine learning systems has become increasingly complex, demanding expertise in distributed computing, specialized frameworks, and systems-level optimization~\cite{bbm24, shb24, vlw25}. Hardware limitations, including restricted memory and bandwidth on accelerators, make large-scale training dependent on intricate strategies like model parallelism, which are challenging to implement and maintain~\cite{yya19,spp19}. These demands are compounded by the frequent need for manual intervention to resolve hardware failures during training~\cite{zrg22}, a burden that academic research groups are too costly to handle.

The gap in access to compute resources is reshaping how research topics are divided between industry and academia in machine learning and foundation models~\cite{shb24, vlw25}. The industry leads the areas where computational power and specialized high performance computing expertise rise while academia focus on lower-compute intensity research and used the open source models from the industry~\cite{bbm24}. 

Due to the wide range of expertise required to develop large-scale foundation models, their technical reports often list hundreds or even thousands of contributors~\cite{team2023gemini, team2024gemini}. In this work, we refer to industry-produced technical reports as \emph{large-scale papers} and academic research publications as \emph{small-scale papers}. 

\subsection{Team Sizes in Foundation Model Research}\label{sec:related_work_team}

Previous studies have attempted to define a threshold to separate large and small research teams. These thresholds have been set either as a fixed number of authors or as a percentage of authors in a dataset. \cite{cro01} first defined hyperauthorship as any paper having more than 100 authors, while another threshold of 200 authors was proposed in~\cite{sta10}. Lower thresholds, such as papers with at least 20 authors, have also been used in~\cite{bv13, sm07}. 
Recent work~\cite{lwl24} proposed a threshold of 26 authors by analyzing author counts using skewed probability distributions and Chebyshev’s inequality. Additionally, percentage-based thresholds, such as classifying the top 3\% or 10\% of papers by author count as large teams, have been proposed in\cite{gaf13} and~\cite{lwl24}. However, these thresholds were derived from datasets without papers containing thousands of authors. Specifically, the largest team size considered in~\cite{lwl24} was only 156 authors while Gemini paper~\cite{team2023gemini} has 1361 authors.

The scale of authorship in recent foundation model publications~\cite{team2023gemini, team2024gemini, grattafiori2024llama3} significantly exceeds previously studied thresholds~\cite{lwl24, bv13}. Traditional definitions and metrics for hyperauthorship~\cite{hir05} are inadequate in scenarios where papers routinely involve hundreds or even thousands of co-authors, as illustrated by the examples of recent foundation model research summarized in Table~\ref{tab:model_stats}.

A large number of co-authors introduces several challenges in research evaluation. While large publishing consortia typically produce research with higher citation impact, attributing contributions to individual authors becomes difficult~\cite{the20}. The Foundation Model Transparency Index~\cite{bkl+23}, which evaluates foundation models from ten leading technology companies, including Llama 2~\cite{touvron2023llama2} and GPT-4~\cite{achiam2023gpt}, reports low transparency regarding data sources, labor contributions, and methodologies, further complicating the assessment of individual author roles.

%% file: 04_problem.tex
\section{Problem Definition}\label{sec:problem_definition}
We model each scholar by a profile
\begin{align*}
    P = \{(c_i,\,a_i)\}_{i=1}^N,
\end{align*}
where $N$ is the total number of the author’s publications, $c_i$ is the citation count of paper $i$ and $a_i$ is the number of co‐authors on that paper. A citation‐based metric is a function
\begin{align*}
    M:\{(c_i,a_i)\}_{i=1}^N \;\longmapsto\; \mathbb{R}_{\ge0}
\end{align*}
that assigns a single nonnegative score summarizing the individual’s overall impact. 
\subsection{Fundamental Properties of Citation Metric}
We require that $M$ satisfy the following \emph{Fundamental Properties of Citation Metric}: 
\begin{itemize}
 \item \emph{Citation Monotonicity.} If any paper’s citation count increases while the number of co-authors remains fixed, the metric does not decrease. If $c_i' > c_i$, then
 \begin{align*}
    M(\{(c_1,a_1), & ~\dots,(c_i',a_i),\dots,(c_N,a_N)\}) \\ 
     \ge & ~ M(\{(c_1,a_1),\dots,(c_i,a_i),\dots,(c_N,a_N)\}).
 \end{align*}
 
 \item \emph{Author Count Penalization.} For a paper with fixed citations, adding more co-authors does not increase the paper's contribution to the final score. If $a_i' > a_i$, then \begin{align*}
    M(\{(c_1,a_1), & ~\dots,(c_i,a_i),\dots,(c_N,a_N)\}) \\ 
     \ge & ~ M(\{(c_1,a_1),\dots,(c_i,a_i'),\dots,(c_N,a_N)\}).
 \end{align*}
 \item \emph{Zero Contribution Baseline.} An author with no citations has zero score. If $c_i = 0$ for all $i\in N$, then $M(\{(c_i,a_i)\}_{i=1}^N) = 0$
\end{itemize}

\subsection{Metric Context}
The primary purpose of this metric is to serve as a practical reference for someone to evaluate researchers in foundation model development. Research in this field spans two distinct styles: large-scale industrial projects, which depend on massive compute and extensive collaboration, and smaller academic efforts, which focus on theoretical advances and innovation. Both styles are critical for success in industry settings. Accordingly, our metric is designed to (1) favor researchers who contribute to both large-scale and small-scale papers, (2) distinguish between researchers with balanced experience across both styles, (3) differentiate those with experience in only one style, and (4) remain transparent and straightforward to compute.

%% file: 05_methods.tex
\section{Methods}\label{sec:methods}
This section introduces the two core metrics we study and the procedures used to tune and evaluate them.  In Section~\ref{sec:method_exph} we formally define the Exponential Fractional $h$-Index and prove that it satisfies our Fundamental Properties of Citation Metrics.  Section~\ref{sec:method_sbci} presents the Scale-Balanced Citation Index (SBCI), including precise definitions of large-scale (L-papers) and small-scale (S-papers) contributions.  In Section~\ref{sec:method_sbci_param} we describe our grid-search procedure for selecting SBCI parameters $(\alpha,\tau,f,g)$ by optimizing a multi-term objective that balances discriminative power, mean and variance alignment, and ranking stability under citation perturbations.  Finally, Section~\ref{sec:method_data_gen} details the synthetic dataset we generate to validate SBCI’s behavior across realistic collaboration scenarios by modeling citation counts, team sizes, publication years, and student types.

\subsection{Exponential fractional \texorpdfstring{$h$}{}-index}\label{sec:method_exph}
In order to explore how co-authorship penalties alter the behavior of the classical $h$-index, we first introduce the \emph{exponential fractional $h$-index} as an intuitive variant of $h$-index.  Unlike the individual $h$-index, which rescales the standard $h$-index by the average number of co-authors, or the fractional $h$-index, which divides each paper’s citations equally among its authors, our exponential fractional version applies a smooth, exponential decay to each paper’s citation count.

\begin{definition}[Exponential fractional $h$-index]\label{def:exp_frac_h}
Let an author’s publication record be 
$\{(c_i,a_i)\}_{i=1}^N$, 
where $c_i$ is the citation count and $a_i$ the number of co‐authors of paper $i$.  
Fix a parameter $\beta>0$.  Define the exponentially fractionalized citation count
\begin{align*}
\tilde c_i = c_i \cdot \exp(-\beta(a_i - 1)).
\end{align*}

Reorder $\{\tilde c_i\}$ in nonincreasing order 
$\tilde c_{1} \ge \tilde c_{2} \ge \dots \ge \tilde c_{N}$.  
Then the \emph{exponential fractional $h$‐index} $h_{\mathrm{EF}}^\beta$ is
\begin{align*}
    h_{\mathrm{EF}}^\beta :=
\max\{h\in\mathbb{N} \mid \tilde c_{h} \ge h\}.
\end{align*}
\end{definition}

We now show that the $h_{\mathrm{EF}}^\beta$ meets the Fundamental Properties of Citation Metric. Let 
\begin{align*}
h_{\mathrm{EF}}^\beta(\{(c_i,a_i)\}^N_{i=1}) 
= \max\{ h\in N \mid \tilde c_{h} \ge h\},
\end{align*}
where $\tilde c_i = c_i e^{-\beta(a_i-1)}.$ We first show that the exponential fractional $h$-index satisfies citation monotonicity. If one citation count $c_j$ increases to $c_j'>c_j$ with all else fixed then $\tilde c_j$ increases to $\tilde c_j' = c_j'\cdot e^{-\beta(a_j-1)}>\tilde c_j$.  No entry decreases in the sorted list, so the largest $h$ with $\tilde c_{(h)}\ge h$ can only stay the same or grow. Then, we show that the this metric satisfies author count penalization.
If one co‐author count $a_j$ increases to $a_j'>a_j$ with all else fixed, then $\tilde c_j' = c_j\cdot e^{-\beta(a_j'-1)} < c_j\cdot e^{-\beta(a_j-1)} = \tilde c_j$. Thus entries in the sorted $\tilde c_{h}$ can only decrease or stay the same, so the maximum valid $h$ cannot increase.
Next, we show that this metric satisfies zero contribution baseline. If $c_i=0$ for all $i$ then every $\tilde c_i=0$. Hence
$\tilde c_{1}=0<1$, so no positive $h$ satisfies $\tilde c_{h}\ge h$, giving $h_{\mathrm{EF}}^\beta=0$.     

\textbf{Remark.} The exponential fractional $h$-index is independent of the SBCI introduced below. We include exponential fractional $h$-index here to illustrate that this intuitive variant of the $h$-index fails to deliver the desired effect in Table~\ref{tab:scholar_metrics}.

\subsection{Scale-Balanced Citation Index (SBCI)}\label{sec:method_sbci}

In this section, we introduce the \emph{Scale‐Balanced Citation Index} (SBCI), a metric that first partitions an author’s scholarly credit into large-scale and small-scale components and then recombines them using a tunable balance weight.

\begin{definition}[Large-scale papers, L-papers]
Let $\tau\in\mathbb{N}$ be an author‐count threshold.  A paper $i$ with $a_i$ co‐authors is called a large‐scale paper (or L‐paper) if $a_i \ge \tau$. We denote the set of all L‐papers in an author’s record by
$L:=\{i\in \{1, \ldots,N\} \mid a_i\ge\tau\}$.  
\end{definition}

\begin{definition}[Small-scale papers, S-papers]
Let $\tau\in\mathbb{N}$ be an author‐count threshold.  A paper $i$ with $a_i$ co‐authors is called a small‐scale paper (or S‐paper) if $a_i < \tau$. We denote the set of all S‐papers in an author’s record by
$S:=\{i\in \{1, \ldots,N\} \mid a_i<\tau\}$.  
\end{definition}

\textbf{Notation.}  For convenience we write \emph{L‐papers} and \emph{S‐papers} instead of $L_{\tau}$-papers and $S_{\tau}$-papers.
In the experiment, we set $\tau=26$ base on~\cite{lwl24}. 

Let a scholar have $N$ total papers, with $c_i$ citations and $a_i$ co-authors for paper $i$.
We first normalze each paper's contribution by assigning a per-paper credit that accounts for author count $w_i:=\frac{c_i}{f(a_i)}$, where $f(a_i)$ grows with $a_i$ (e.g., $f(a_i)=a_i$ for simple fractional credit, or $f(a_i)=\sqrt{a_i}$ to avoid over-penalizing large teams). Then, we define $W_L := \sum_{i\in L} w_i$ and $W_S := \sum_{i\in S} w_i$, where $L$ and $S$ are the sets of large-scale and small-scale papers. Instead of letting one dominate, we combine them as $M = \alpha \cdot g(W_L) + (1-\alpha) \cdot g(W_S),$ where $0 \leq \alpha \leq 1$ is a weight that can be fixed, tuned, or learned from real-time data, $g(\cdot)$ could be identity, log, or a capped function to avoid citation explosions. Formally, we define the SBCI as follow. 

\begin{definition}[Scale‐Balanced Citation Index (SBCI)]
Let an author’s publication record be 
$P = \{(c_i,a_i)\}_{i=1}^N$, 
where $c_i\in\mathbb{N}$ is the citation count and $a_i\in\mathbb{N}$ is the number of co‐authors of paper $i$. Given the following: 

\begin{itemize}
  \item A nondecreasing author count penalty function $f:\mathbb{N}\to\mathbb{R}^+$,  
  \item A nondecreasing citation normalization function $g:\mathbb{R}^+\to\mathbb{R}^+$,  
  \item A balance weight $\alpha\in[0,1]$,  
  \item An author‐count threshold $\tau\in\mathbb{N}$ to distinguish large scale and small scale papers.
\end{itemize}
Define for each paper $i$, $w_i := \frac{c_i}{f(a_i)},$
and let $L = \{i \mid a_i \ge \tau\}$ and $S = \{i \mid a_i < \tau\}$ be the large-scale and small scale paper sets. Then the \emph{Scale‐Balanced Citation Index} is
\begin{align*}
    \mathrm{SBCI}_{\alpha,f,g,\tau}(P) := 
\alpha\cdot g(\sum_{i\in L} w_i) + (1-\alpha)\cdot g(\sum_{i\in S} w_i).
\end{align*}

\end{definition}


We now verify that SBCI satisfies each of the Fundamental Properties of a citation metric. Suppose a single citation count $c_j$ increases to $c_j'>c_j$ while all other $(c_i,a_i)$ remain unchanged.  Then the per‐paper credit
 \begin{align*}
w_j  = \frac{c_j}{f(a_j)}
\rightarrow
w_j'  = \frac{c_j'}{f(a_j)}  >  w_j.
 \end{align*}
Consequently, the aggregate credit for either the large‐scale set $W_L$ (if $j\in L$) or the small‐scale set $W_S$ (if $j\in S$) strictly increases.  Because $g$ is nondecreasing, we have $g(W_L') \ge g(W_L)$ and $g(W_S') \ge g(W_S),$
so $\mathrm{SBCI}'  = \alpha \cdot g(W_L') + (1-\alpha) \cdot g(W_S') \ge \alpha \cdot g(W_L) + (1-\alpha) \cdot g(W_S) = \mathrm{SBCI}$. Suppose the co‐author count $a_j$ increases to $a_j' > a_j$ with $c_j$ fixed.  Since $f$ is nondecreasing,
$f(a_j')  \ge  f(a_j) \rightarrow w_j'  = \frac{c_j}{f(a_j')}  \le \frac{c_j}{f(a_j)}  =  w_j.$ Thus the affected aggregate $W_L$ or $W_S$ cannot increase.  Applying the same nondecreasing $g$ argument shows that $\mathrm{SBCI}'\le \mathrm{SBCI}$, so adding co‐authors cannot raise an author’s SBCI score. Finally, if every citation count $c_i = 0$, then each per‐paper credit $w_i = 0$, giving $W_L = W_S = 0$.  Therefore $\mathrm{SBCI}  = \alpha g(0) + (1-\alpha) g(0)  =  g(0) = 0$, assuming $g(0)=0$, so an author with no citations receives a score of zero.

\subsection{Parameter Selection for SBCI}\label{sec:method_sbci_param}

Let $M_\theta(i)$ denote the SBCI score for researcher $i$ computed with parameters 
$\theta=(\alpha,\tau,f,g)$, and let $W_{L,i}$ and $W_{S,i}$ be their aggregated, 
$\alpha$- and $(1-\alpha)$-weighted credits from large‑scale and small‑scale papers,
respectively. We choose $\theta$ by maximizing an objective that trades off three criteria:
\begin{equation}\label{eq:obj}
\begin{aligned}
\theta^* = \arg\max_\theta \{
  &\underbrace{\alpha \cdot \mathrm{Var}(g(W_{L, i}))+(1-\alpha) \cdot \mathrm{Var}(g(W_{S, i}))}_{\mathrm{Discriminative\ Power}}\\
  &-\lambda_{1}\underbrace{|\alpha \cdot \mu(g(W_{L, i}))+(1-\alpha) \cdot \mu(g(W_{S, i}))|}_{\mathrm{Mean\ Balance}}\\
  & -\lambda_{2}\underbrace{|\alpha \cdot \mathrm{Var}(g(W_{L, i}))-(1-\alpha) \cdot \mathrm{Var}(g(W_{S, i}))|}_{\mathrm{Variance\ Balance}}\\
  & -\lambda_{3}\underbrace{\frac{1}{N}\sum_i|\mathrm{rank}_\theta(i) - \mathrm{rank}_\theta^\epsilon(i)|}_{\mathrm{Stability\ Penalty}}
\}.
\end{aligned}
\end{equation}
In Equation~(\ref{eq:obj}), we let
\begin{align*}
\mu(g(W_{L,i}))
 = \frac{1}{N}\sum_{i=1}^N g(W_{L,i}),
\end{align*}
and similarly for $\mu(g(W_{S,i}))$; here $\mu$ denotes the empirical mean over all $N$ researchers. The optimization seeks parameters $\theta=(\alpha,\tau,f,g)$ that maximize a composite objective of four components. The first component, \emph{Discriminative Power},
\begin{align*}
\alpha \cdot \mathrm{Var}_i(g(W_{L,i})) + (1-\alpha) \cdot \mathrm{Var}_i(g(W_{S,i})),
\end{align*}
promotes high variability in both large-scale and small-scale credits, ensuring SBCI can distinguish authors across collaboration scales. The second component, \emph{Mean Balance},
\begin{align*}
-\lambda_{1} |\alpha \cdot \mu(g(W_{L,i}))+(1-\alpha) \cdot \mu(g(W_{S,i}))|,
\end{align*}
penalizes disproportionate average contributions, guiding $\alpha$ toward equalizing mean impacts from each scale. The third component, \emph{Variance Balance},
\begin{align*}
-\lambda_{2} |\alpha \cdot \mathrm{Var}_i(g(W_{L,i}))-(1-\alpha) \cdot \mathrm{Var}_i(g(W_{S,i}))|,
\end{align*}
further constrains the metric to avoid dominance by either scale. Finally, the \emph{Stability Penalty},
\begin{align*}
-\lambda_{3} \frac{1}{N}\sum_{i=1}^N|\mathrm{rank}_\theta(i)-\mathrm{rank}_\theta^\epsilon(i)|,
\end{align*}
ensures robustness by limiting rank fluctuations under small citation perturbations $\epsilon$. Let $\theta^*=(\alpha^*,\tau^*,f^*,g^*)$ denote the parameter tuple that maximizes the objective function. It provides an SBCI metric with strong discriminative power, balanced scale contributions, and stable author rankings.

\subsection{Synthetic Dataset for Validation}\label{sec:method_data_gen}

Since no public dataset captures the joint distribution of team sizes, citation counts, and publication years for machine learning PhD students, we generate a synthetic dataset to evaluate and tune SBCI. The dataset models 1000 simulated PhD students over a five-year period (2020–2024), where each student publishes between 4 and 10 papers during their program.

We model citation counts per paper using a log-normal distribution:
\begin{align*}
    c_i \sim \mathrm{LogNormal}(\mu_c, \sigma_c),
\end{align*}

with parameters $(\mu_c,\sigma_c) = (2.5,1.2)$ to reflect the heavy-tailed nature of computer science citation patterns. Team sizes are sampled from a mixture distribution to capture both small academic teams and large industrial collaborations:
\begin{align*}
    a_i \sim p \cdot \mathrm{Poisson}(\lambda_s) + (1-p) \cdot (a_{\min} + \mathrm{Pareto}(\alpha)),
\end{align*}

with $\lambda_s=4$, $\alpha=1$, $a_{\min}=26$ as the threshold for large-scale teams~\cite{lwl24}, and $p$ depending on the student type. In constructing the dataset, we categorize students into three groups to reflect distinct research profiles. The first group consists of 800 students who publish exclusively in small-team settings, with all of their papers sampled from the Poisson component ($p=1$). The second group contains 100 students who contribute to both small- and large-scale papers, using a mixture probability $p=0.7$ to determine team size per paper. The final group consists of 100 students who publish exclusively in large-team settings, with all of their papers drawn from the Pareto component ($p=1$). This last group represents individuals whose contributions often focus on software engineering, coordination, or infrastructure tasks typical of industrial-scale collaborations. 

To capture real-world dynamics, citations are scaled by team size and publication year. Large-team papers accrue disproportionately more citations, with a scaling factor:
\begin{align*}
    c_i \gets c_i \cdot 
    \begin{cases}
    1 + 1.5 \log(1+a_i), & a_i \geq 26, \\
    1 + 0.5 \log(1+a_i), & a_i < 26,
    \end{cases}
\end{align*}

and recent papers gain citations more rapidly using an age-based adjustment:
\begin{align*}
    c_i \gets c_i / (2025 - y_i + 1)^{-0.85},
\end{align*}
where $y_i$ is the publication year. To avoid unrealistic outliers, all citation counts are capped at 5,000. The final dataset, which includes each paper’s year, number of co-authors, and total citations, is used for tuning the SBCI parameters and for validation experiments.

Each student group shares the same citation and age-scaling mechanisms, but their team-size distributions reflect their distinct research styles. This stratification allows SBCI to be evaluated on its ability to reward balanced contributors, differentiate between researchers who specialize exclusively in one style, and avoid collapsing scores for hyperauthored categories.

%% file: 06_experiment.tex
\section{Experiment}\label{sec:experiment}
In this section, we first tune SBCI’s hyperparameters via an exhaustive grid search on a synthetic dataset of 1,000 ML PhD students by optimizing discriminative power, mean/variance balance, and ranking stability under citation perturbations (Section~\ref{sec:exp_sbci_param}).  We then demonstrate SBCI alongside traditional indices in a controlled six‐candidate case study (Section~\ref{sec:exp_comparison}).  Finally, we compare alternative SBCI configurations and recommend $\alpha=0.6$, $f(a)=\sqrt{a}$, and $g(w)=\log(1+w)$ as our default parameter set.

\subsection{Determining SBCI Parameters}\label{sec:exp_sbci_param}
To identify a robust range for the balance weight $\alpha$, we conducted an exhaustive grid search over $\alpha\in\{0.0,0.2,0.4,0.6,0.8,1.0\}$, the author‐penalty functions $f(a)\in\{a,\sqrt{a}\}$, and the citation‐normalization functions $g(w)\in\{w,\log(1+w)\}$, while holding $\tau=26$, $\lambda_2=1.0$, and $\lambda_3=1.0$ fixed.  For each $(\alpha,f,g)$ combination we evaluated the full objective in Equation~(\ref{eq:obj}), including the stability term under a ten‐citation perturbation ($\epsilon=10$) of every paper.

Table~\ref{tab:alpha_fg_results} presents the resulting scores.  When $f(a)=\sqrt{a}$ and $g(w)=\log(1+w)$, the objective peaks at $\alpha=0.6$ (value $-0.32$), whereas for $f(a)=\sqrt{a}$ and $g(w)=w$ the maximum occurs at $\alpha=0.8$ (value $241.67$).  Similarly, under the linear penalty $f(a)=a$, both normalization choices favor $\alpha=0.8$ ($-0.32$ for $\log$, $-3.43$ for identity).  In most cases the extreme endpoints $\alpha=0.0$ and $\alpha=1.0$ yield the lowest scores, confirming that neither pure reliance on small-scale nor on large-scale contributions suffices.  Across all four $(f,g)$ blocks, mid‐range values $\alpha=0.6$ or $\alpha=0.8$ consistently maximize the objective. This result demonstrates that a balanced weighting is essential for discriminative power and stability.

\begin{table}[!ht]
\centering
\begin{tabular}{r l l r}
\toprule
$\alpha$ & $f(a)$ & $g(w)$ & Objective \\
\midrule
0.0 & $\sqrt{a}$ & $\mathrm{log1p}(w)$ & -3.91 \\
0.2 & $\sqrt{a}$ & $\mathrm{log1p}(w)$ & -2.55 \\
0.4 & $\sqrt{a}$ & $\mathrm{log1p}(w)$ & -1.40 \\
0.6 & $\sqrt{a}$ & $\mathrm{log1p}(w)$ & \textbf{-0.32 }\\
0.8 & $\sqrt{a}$ & $\mathrm{log1p}(w)$ & -1.23 \\
1.0 & $\sqrt{a}$ & $\mathrm{log1p}(w)$ & -2.58 \\
\midrule
0.0 & $\sqrt{a}$ & $w$ & -62.22 \\
0.2 & $\sqrt{a}$ & $w$ & -374.75 \\
0.4 & $\sqrt{a}$ & $w$ & -427.70 \\
0.6 & $\sqrt{a}$ & $w$ & -221.05 \\
0.8 & $\sqrt{a}$ & $w$ & \textbf{241.67} \\
1.0 & $\sqrt{a}$ & $w$ & -18.08 \\
\midrule
0.0 & $a$ & $\mathrm{log1p}(w)$ & -3.40 \\
0.2 & $a$ & $\mathrm{log1p}(w)$ & -2.49 \\
0.4 & $a$ & $\mathrm{log1p}(w)$ & -1.48 \\
0.6 & $a$ & $\mathrm{log1p}(w)$ & -0.62 \\
0.8 & $a$ & $\mathrm{log1p}(w)$ & \textbf{-0.32} \\
1.0 & $a$ & $\mathrm{log1p}(w)$ & -1.22 \\
\midrule
0.0 & $a$ & $w$ & -36.98 \\
0.2 & $a$ & $w$ & -196.67 \\
0.4 & $a$ & $w$ & -269.00 \\
0.6 & $a$ & $w$ & -253.90 \\
0.8 & $a$ & $w$ & -151.53 \\
1.0 & $a$ & $w$ & \textbf{-3.43} \\
\bottomrule
\end{tabular}
\caption{Grid‐search objective scores for each $\alpha$, $f$, $g$ combination.}
\label{tab:alpha_fg_results}
\end{table}

\subsection{Metrics under Foundation Model Collaboration}\label{sec:exp_comparison}

\begin{figure*}[!ht]
    \centering
    \includegraphics[width=0.7\linewidth]{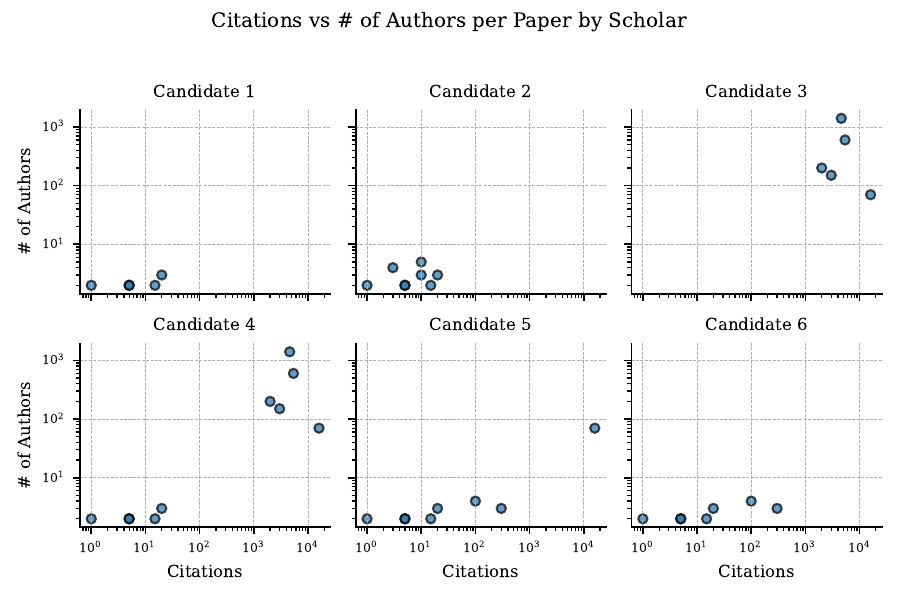}
    \caption{Log–log scatter of citations per paper versus number of authors of six PhD-student in the job market. }
    \label{fig:candidates}
\end{figure*}

\begin{table*}[!ht]
    \centering 
        
        \begin{tabular}{lcccccccccc}
            \toprule
            Cand. & L-scale & S-scale & Total & Total & $h$ & $h_{\rm I}$ & $h_{\rm frac}$ & $g$ & $h_{\rm exp}$ & SBCI\\
            & papers & papers & Papers & Citations & index & index & index & index & index & index \\
            \midrule
1 & 0 & 5 & 5 & 46 & 4 & 1.78 & 2 & 5 & 2 & 1.37 \\
2 & 0 & 8 & 8 & 69 & 5 & 1.67 & 3 & 8 & 2 & 1.50 \\
3 & 5 & 0 & 5 & 31000 & 5 & 0.01 & 4 & 5 & 0 & 4.73 \\
4 & 5 & 5 & 10 & \textbf{31046} & \textbf{7} & 0.02 & \textbf{6} & \textbf{10} & 2 & 6.10 \\
5 & 1 & 7 & 8 & 16446 & 5 & 0.30 & 5 & 8 & \textbf{3} & \textbf{6.75} \\
6 & 0 & 7 & 7 & 446 & 5 & \textbf{1.79} & 4 & 7 & \textbf{3} & 2.22 \\
            \bottomrule
        \end{tabular}
    
    \caption{Case‐study evaluation of six PhD‐student candidates using simulated (fake) citation and authorship data under traditional and normalized metrics. L-scale is Large-scale papers and S-scale represent the Small-scale papers. The SBCI parameters are $\alpha=0.6$, $f(a)=\sqrt{a}$, $g(w)=\log(1+w)$, and $\tau=26$. A more detailed SBCI score is in Table~\ref{tab:sbci_results}.}
    \label{tab:scholar_metrics}
\end{table*}

In this section, we present a toy “job interview’’ scenario with six PhD‐student candidates. As shown in Figure~\ref{fig:candidates}, each candidate’s profile is summarized by their per‐paper citation counts and co‐author numbers:
 
\begin{itemize}
  \item \emph{Candidate 1:} Five modest contributions from small research teams, with citations ranging from 1 to 20.
  \item \emph{Candidate 2:} Eight publications by small–to–medium teams, each cited 3–20 times. Notices that this set includes all five of Candidate 1’s publications for direct comparison.
  \item \emph{Candidate 3:} Five high-impact consortium papers from very large collaborations, each cited thousands of times.
  \item \emph{Candidate 4:} Five large scale and five small scales papers, with citation counts ranging from 1 to 16 000. Notices that this set includes all five publications of candidate 1 and five publications of candidate 3.  
  \item \emph{Candidate 5:} Eight diverse contributions including one blockbuster hyper-cited paper and several moderate small-team works, citations between 1 and 16 000. Notices that this set includes all five publications of candidate 1 for direct comparison.
  \item \emph{Candidate 6:} Seven solid contributions from small research groups, with citations ranging from 1 to 300. Notice that this set contains the same small scale paper as candidate 5. 
\end{itemize}
Assume a committee must select one candidate purely on the basis of Figure~\ref{fig:candidates} and Table~\ref{tab:scholar_metrics}, with no access to actual publications or personal information.  If they use the raw $h$-index as their sole criterion, they would rank:
\begin{align*}
    \mathrm{Candidate~4 }(h=7)~\succ & ~ \mathrm{Candidates~2,~3,~5,~6 }(h=5) \\ 
    \succ & ~ \mathrm{Candidate 1 }(h=4).
\end{align*}
By contrast, if they rely on the individual $h$-index $h_{\rm I}$, which discounts for co-authorship, the ordering flips:
\begin{align*}
    \mathrm{Candidate~6}(h_I=1.79)~\succ & ~ \mathrm{Candidate~1}(h_I=1.78) \\ 
    \succ & ~ \mathrm{Candidate~2}(h_I=1.67)\\ 
    \succ & ~ \dots
\end{align*}

If they instead use the fractional $h$-index $h_{\rm frac}$, which shares each paper’s citations equally among its co‐authors, they obtain:
\begin{align*}
    \mathrm{Candidate~4 }(h_{\rm frac}=6)~ \succ & ~ \mathrm{Candidate~5 }(h_{\rm frac}=5) \\
    \succ & ~ \mathrm{Candidates~3,6}(h_{\rm frac}=4) \\ 
    \succ & ~ \dots
\end{align*}

Using the $g$-index again favors consortium-heavy records, selecting Candidate 4 $(g=10)$ at the top. The exponential fractional $h$-index, $h_{\mathrm{exp}}$, also performs poorly. All students receive indistinguishable scores of 0, 2, or 3. Moreover, due to the exponential weighting, Candidate 3 receives no credit for the publications.

Thus:
\begin{itemize}
  \item \emph{$h$-index}: Candidate 4 with massive consortium papers wins.
  \item \emph{Individual $h_{\rm I}$-index}: Candidate 6 with small-team consistency wins.
  \item \emph{Fractional $h_{\rm frac}$-index}: Candidate 4 again wins, but Candidate 5 rises to second place.
\end{itemize}

This stark contradiction shows that traditional metrics can drive \emph{opposite} suggestions depending on which index is chosen.

\begin{table}[!ht]

\centering
\begin{tabular}{lrrrr}
\toprule
Cand. & SBCI & SBCI & SBCI & SBCI \\
 & $.6/\sqrt{(\cdot)}/\log(\cdot)$ & $.8/\sqrt{(\cdot)}/\mathrm{id}$ & $.8/\mathrm{id}/\mathrm{\log(\cdot)}$ & $0.8/\mathrm{id}/\mathrm{id}$\\
\midrule
1 & 1.37 & 5.99 & 0.61 & 3.93 \\
2 & 1.50 & 8.34 & 0.66 & 5.15 \\
3 & 4.73 & 2113.70 & 4.48 & 216.69 \\
4 & 6.10 & 2119.69 & 5.09 & 220.62 \\
5 & 6.75 & 1580.52 & 5.35 & 211.79 \\
6 & 2.22 & 50.63 & 1.00 & 28.93 \\
\bottomrule
\end{tabular}
\caption{SBCI scores for six candidate profiles under four parameter settings.  The notation “$\alpha/f(a)/g(w)$” in each column header indicates the weight $\alpha$, the author‐penalty function $f$, and the citation‐normalization function $g$. $\mathrm{id}$ refers to $f(x)=x$. For example, “0.6/$\sqrt{a}$/$\log$” denotes $\alpha=0.6$, $f(a)=\sqrt{a}$, and $g(w)=\log(1+w)$.}
\label{tab:sbci_results}
\end{table}

Table~\ref{tab:sbci_results} reports SBCI scores for each of the six candidate profiles under the four optimized $(\alpha,f,g)$ settings. Although each column corresponds to the maximum objective achieved for its own $f$ and $g$, the resulting score scales differ dramatically: for example, using $f(a)=\sqrt{a}$ and $g(w)=\mathrm{id}$ produces values in the thousands for hyper-authored candidates, whereas $f(a)=\mathrm{id}$ and $g(w)=\log(1+w)$ keeps all scores below 6. Such disparities hinder direct comparison across columns and complicate interpretation.

We therefore adopt $\alpha=0.6$, $f(a)=\sqrt{(\cdot)}$, and $g(w)=\log{1+w}$ as our default.  This configuration yields moderate SBCI values (e.g.\ 6.75 for Candidate 5), maintains clear separation between small- and large-scale profiles, and preserves rank stability under small citation perturbations.  Importantly, the choice of $\alpha$, $f$, and $g$ remains flexible: as the balance between small-team and large-team contributions shifts over time, or as industry hiring priorities evolve, one can re-run the same grid-search procedure to retune these parameters and adapt SBCI to new requirements.  

%% file: 07_discussion.tex
\section{Discussion}\label{sec:discussion}

Our experiments demonstrate that SBCI provides a practical, interpretable signal of individual scholarly impact in the era of hyper‐authored foundation models.  We emphasize, however, that SBCI is intended as one component of a holistic evaluation process rather than a standalone index.  Real‐world recruitment involves interviews, reference checks, domain expertise, and alignment with organizational goals, all of which must be weighed alongside any citation‐based metric.

We also observe that SBCI may assign relatively high scores to researchers whose records consist exclusively of large‐scale industrial reports.  In many cases, these contributors perform vital software‐engineering, infrastructure, or project‐management roles rather than core scientific leadership.  We view this not as a flaw but as a feature: SBCI faithfully reflects credit‐sharing across all roles in a collaboration in foundation model research, and it is up to user (e.g., hiring team or committee) to interpret those scores in context.  Indeed, teams require diverse skill sets—both research innovation and engineering execution—to succeed at scale.

To support more fine‐grained assessment, we advocate for richer contributorship disclosures in large consortium publications.  Adopting standardized taxonomies such as the CRediT system or organizing authors into functional groups (e.g., “Model Architecture,” “Data Curation,” “Infrastructure”) would enable SBCI to weight contributions more precisely. Future work could integrate such metadata directly into the penalty function $f(a)$ or the normalization function $g(w)$, further balancing recognition between intellectual leadership and technical support.

Finally, SBCI’s reliance on citation and co‐author counts means it inherits the well‐known limitations of bibliometric data, including publication delays, field‐specific citation practices, and potential gaming.  We recommend that organizations use SBCI in conjunction with qualitative reviews, direct inspect of publication content, and other impact indicators (e.g., software releases, patent filings) to ensure a fair, comprehensive evaluation of candidate contributions.

%% file: 08_conclusion.tex
\section{Conclusion}\label{sec:conslusion}

In this work, we have identified a critical gap in traditional citation metrics when applied to the era of hyper‐authored foundation model research.  We introduced the Scale‐Balanced Citation Index (SBCI) and proved that they satisfy fundamental axioms of citation impact, including monotonicity, co‐author penalization, and diminishing marginal reward. Through a synthetic dataset reflecting real‐world citation, team‐size, and publication‐year patterns, we demonstrated how SBCI can be tuned via a composite objective to balance large‐ and small‐scale contributions while maintaining ranking stability under citation perturbations.  A case study with six candidate profiles further illustrated SBCI’s ability to discriminate among researchers whose traditional $h$, $g$, or fractional $h$ indices would collapse to similar values.

Moving forward, SBCI provides a flexible, interpretable signal that can inform, but not replace, holistic evaluation processes in hiring, promotion, and funding decisions.  Its parameters $(\alpha,\tau,f,g)$ can be re‐optimized as collaboration patterns evolve or as organizational priorities change.  We encourage future work to integrate detailed contributorship metadata, to validate SBCI against real distribution, and to extend our framework to other domains with emerging hyperauthorship trends. By combining SBCI with qualitative peer review and alternative impact measures, institutions can more fairly and accurately recognize individual contributions in large‐scale foundation model research collaborations.